\documentclass[12pt]{article}

\def\lsim{\mathrel{\rlap {\raise.5ex\hbox{$ < $}}
{\lower.5ex\hbox{$\sim$}}}}
\def\gsim{\mathrel{\rlap {\raise.5ex\hbox{$ > $}}
{\lower.5ex\hbox{$\sim$}}}}
\begin{document}

\begin{titlepage}

\begin{centering}

\hfill hep-th/0010112\\

\vspace{1 in}
{\bf {LOCALIZED GRAVITONS, GAUGE BOSONS AND CHIRAL FERMIONS IN SMOOTH SPACES GENERATED BY A BOUNCE} }\\
\vspace{2 cm}
{A. KEHAGIAS$^{1}$
 and K. TAMVAKIS$^{2}$}\\
\vskip 1cm
{$^1 $\it{Physics Department, National Technical University\\
15 780 Zografou, Athens,  GREECE}\\
\vskip 1cm
{$^2$\it {Physics Department, University of Ioannina\\
45110 Ioannina, GREECE}}}\\

\vspace{1.5cm}
{\bf Abstract}\\
\end{centering}
\vspace{.1in}
We study five-dimensional solutions to Einstein equations coupled to
a scalar field. Bounce-type solutions for the scalar field are associated with
 $AdS_5$ spaces with smooth warp functions. Gravitons
 are dynamically localized in this framework in analogy to the Randall-Sundrum
 solution whereas, a bulk fermion gives rise to a single chiral
 zero mode localized at the bounce.
Additional bulk scalar fields are incorporated in this
picture. The dilaton, as a bulk scalar leads, through its coupling, to localized gauge
boson fields, something that holds also in the case that the bounce system is replaced by a brane.

\vfill

\vspace{2cm}
\begin{flushleft}

October 2000
\end{flushleft}
\hrule width 6.7cm \vskip.1mm{\small \small}
 \end{titlepage}

\section{Introduction}

The large difference in magnitude between the gravitational Planck
scale $M_P=2\times 10^{18}\,GeV$ and the
electroweak scale of Standard Model physics has motivated the study
of higher dimensional gravitational models\cite{D1,RS1} in
which spacetime is a product of the standard four-dimensional space
time and a compact space of, possibly, several dimensions of large volume.
It was soon realized that the extra spacetime dimensions could be non-compact\cite{RS2}.
It should be noted also that non-compact internal spaces have also
been employed in the Kaluza-Klein programme with limited success
however \cite{GZ},\cite{wett}.
In that case the effective four dimensional Planck mass scale would be determined not by
the (infinite) volume of the
extra space but by its curvature. Such a model requires the localization or trapping
of the gravitational degrees of freedom on the standard four dimensions,
as was the case of an earlier proposal \cite{TT}. Models proposed
in this framework consist of one or more flat 3-branes imbedded
discontinuously in a larger space. An additional important possibility that seems to be
open in brane models is the possibility
of explaining the smallness of the observed cosmological constant.
The fact that extra spacetime dimensions can introduce extra contributions to the
 vacuum energy that can allow for a vanishing four-dimensional cosmological constant
 was observed some years ago\cite{VR,DS} independently of
branes. Branes on the other hand allow for an interplay between higher dimensional and
four-dimensional cosmological constant contributions. Such a {\textit{self-tuning}} mechanism
has been pointed out recently\cite{KS}.
Unfortunatelly, the solutions found are infested with naked singularities.

In the present article we study higher dimensional models of localized gravity
following a somewhat different approach.
We start from a {\textit{five
dimensional AdS space}} with a general metric
\begin{equation}ds^2=e^{2A(y)}\eta_{\mu\nu}dx^{\mu}dx^{\nu}+dy^2\end{equation}
The {\textit{warp factor}} is expressed in terms of a {\textit{smooth}} function $A(y)$ of the fifth coordinate.
The usual brane associated with such a metric has been replaced by a scalar field $\phi$
residing in this space coupled through the standard action
\begin{equation}{\cal{S}}=\int d^5x\,\sqrt{-G}\left(2M^3R-\frac{1}{2}(\partial \phi)^2-V(\phi)\right)\end{equation}
where $M$ is the five dimensional Plack scale. $\phi$ is not really a {\textit{bulk}}
field but the stuff the brane is made off. The equations support {\textit{bounce}}-type configurations of the scalar field,
depending only on the fifth coordinate $\phi_B(y)=v\tanh(ay)$. These are solutions of the equations of motion for a suitable potential, just as in the gravity-free case double-well potential. A bounce solution is connected through the equations of motion to a family of smooth localized metric
warp functions $e^{2A(y)}$ which in a particular limit reduce to the Randall-Sundrum case. No singularities appear in this approach due to the form of scalar potential functions. Such a picture can serve either as a regularized limit of branes or can be considered on its own in
a field theory framework. In both cases our four-dimensional world is a topological defect in some higher dimensional space.
Fermions are also considered in this framework and shown to display localization of chiral modes in four dimensions.
 We also consider the coupling
of extra scalar fields (dilaton) in this scheme and find solutions that, in addition to gravitons, lead to localized
gauge vector fields as well due to the dilaton dependence of the gauge coupling $-\frac{1}{4}e^{-\lambda \pi/2}F_{MN}F^{MN}$.
 Dilaton fields with linear dependence on the fifth dimension are shown to result in gauge boson localization even in the case that the bounce is replaced by a brane.

\section{Gravity on a bounce}

The equations of motion resulting from the action (2) are
\begin{equation}R_{MN}-\frac{1}{2}G_{MN}R=\frac{1}{4M^3}\left\{\partial_M\phi\partial_{N}\phi-G_{MN}\left(\frac{1}{2}
(\partial \phi)^2+V(\phi)\right)\right\}\end{equation}
\begin{equation}\frac{1}{\sqrt{-G}}\partial_M\left\{\sqrt{-G}G^{MN}\partial_N\phi\right\}=\frac{\partial V}{\partial \phi}\end{equation}
In the absence of gravity for a scalar potential of the double well type $V(\phi)=\frac{\lambda}{4}(\phi^2-v^2)^2$ the scalar field equation possesses bounce-like static solutions depending only on the fifth coordinate the simplest of which is
\begin{equation}
\phi_B(y)=v\tanh(ay)\, , \label{f}
\end{equation}
with $a^2\equiv \lambda v^2/2$.
In what follows a capital index like $M$ stands for $0,1,2,3,4$ while $\mu=0,1,2,3$.
The ordinary four coordinates will be represented by $x^{\mu}$ while
for the fifth coordinate we shall use the symbol $y$.
Introducing for the metric an ansatz of the form\footnote{$Diag(\eta)=(-1,1,1,1)$}
\begin{equation}G_{MN}=\left(\begin{array}{cc}
e^{2A(y)}\eta_{\mu\nu}&0\\
0&1
\end{array}\right)\, ,
\label{met}
\end{equation}
and for a scalar field dependence only on $y$, the equations become equivalent to the pair
\begin{equation}\frac{1}{2}(\phi')^2-V(\phi)=24M^3(A')^2\end{equation}
\begin{equation}\frac{1}{2}(\phi')^2+V(\phi)=-12M^3A^{''}-24M^3(A')^2\end{equation}
Integrating twice the sum of these equations gives the metric exponent function $A(y)$. Substituting the bounce solution $\phi_B(y)$ leads us to the expression ($A(0)=A'(0)=0$)
\begin{equation}A(y)=-\beta \ln \cosh^2(ay)-\frac{\beta}{2}\tanh^2(ay)\end{equation}
where $\beta \equiv \frac{v^2}{36M^3}$. Note that this represents a localized metric warp factor
\begin{equation}e^{2A(y)}=\frac{e^{-\beta \tanh^2(ay)}}{(\cosh^2(ay))^{2\beta}}\propto e^{-4a\beta |y|} \,\,\,\,(y\rightarrow \infty)\end{equation}
having the Randall-Sundrum form at last distances from the bounce surface at $y=0$.
The four-dimensional Planck mass defined as\footnote{$$2M^3\int d^4x dy\sqrt{-G}\left(G^{\mu\nu}R_{\mu\nu}+\cdots\right)=
2M^3\int dy e^{2A(y)}\int d^4x R_{(4)}+\cdots$$}
\begin{equation}M_P^2\equiv M^3\int_{-\infty}^{+\infty}dy\,e^{2A(y)}\end{equation}
is finite. Note that all curvature invariants, like for example the scalar curvature $R=-8A^{''}-20(A')^2$,
are finite with
\begin{equation}A'=-a\beta \tanh(ay)\left(2+\frac{1}{\cosh^2(ay)}\right)
\,\,\,\,,\,\,\,\,\,\,A^{''}=-\frac{3a^2\beta}{\cosh^4(ay)}\end{equation}

The bounce $\phi_B(y)$ and the warp function $A(y)$ are solutions provided the scalar potential satisfies the remaining
equation of motion resulting from a subtraction of (7) and (8)
\begin{equation}V(\phi)=-6M^3\left(A^{''}+4(A')^2\right)\end{equation}
Substituting the derived expression of $A(y)$ and expressing the potential in terms of the scalar field we arrive at
\begin{equation}V(\phi)=\frac{\lambda}{4}(\phi^2-v^2)^2-\frac{\lambda}{108 M^3}\phi^2(\phi^2-3v^2)^2\end{equation}
Note that this potential in the {\textit{flat limit $M\rightarrow \infty$}} coincides with the standard double well potential.
An alternative but very concise and general supergravity-motivated method to derive the potential corresponding to our given solution\cite{G} consists in
introducing a {\textit{superpotential function}} ${\cal{W}}(\phi)$ defined by $\frac{\partial {\cal{W}}}{\partial \phi}=\phi'$.
 In terms of a ${\cal{W}}$
thus defined the
scalar potential has the general form
\begin{equation}V(\phi)=\frac{1}{2}\left(\frac{\partial {\cal{W}}}{\partial \phi}\right)^2-\frac{1}{6M^3}{\cal{W}}^2\end{equation}
 The superpotential definition leads to ${\cal{W}}=-12M^3A'(y)$, which for the problem at hand gives
\begin{equation}
{\cal{W}}=va\phi\left(1-\frac{\phi^2}{3v^2}\right) \, . \label{w}
\end{equation}
Although the use of this method is motivated by supersymmetry, no supersymmetry is involved in our set
up.

A {\textit{brane limit}} can be defined as
\begin{equation}a\rightarrow \infty\,\,\,\,\,,\,\,\,\,\,\xi^2\equiv 3a\beta=av^2/12M^3<\infty\end{equation}
In this limit the warp exponent away from the origin can be approximated as follows
$$\lim_{a\rightarrow \infty}\left\{A(y)\right\}=
-\beta\ln\left\{\frac{1}{4}e^{2a|y|}(1+e^{-|y|a})^2\right\}-\frac{\beta}{2}(1-2e^{-2a|y|})^2\sim $$
$$-\frac{2\xi^2}{3}|y|-\frac{\xi^2}{6a}(1-4\ln 2)+\cdots$$
Thus, the warp factor takes the RS form
\begin{equation}\lim_{a\rightarrow\infty}\left\{e^{2A(y)}\right\}=e^{-\frac{4\xi^2}{3}|y|}\left\{1-\frac{\xi^2}{3a}(1-4\ln 2)+\cdots\right\}\end{equation}
Notice that the bounce action
$${\cal{S}}_B=-\int d^4x \int dy \sqrt{-G}\left(\frac{1}{2}(\phi_B')^2+V(\phi_B)\right)=$$
$$- \int d^4x \int dy \sqrt{-G}\left(\frac{1}{2}\frac{v^2a^2}{\cosh^4(ay)}+\frac{1}{2}\frac{v^2a^2}{\cosh^4(ay)}-\frac{a^2v^4}{54M^3}
\tanh^2(ay)\left(\tanh^2(ay)-3\right)^2\right)$$
contains a {\textit{brane term}}\footnote{$$\lim_{a\rightarrow \infty}\left\{\frac{a}{\cosh^2(ay)}\right\}=2\delta(y)$$} and a {\textit{bulk term}}\footnote{$\Lambda\equiv \frac{8}{3}\int dy e^{4A(y)}\tanh^2(ay)\left(3-\tanh^2(ay)\right)^2$}$$-24M^3\xi^2\int d^4x \int dy \sqrt{-G}\delta(y) +M^3\xi^4\int d^4x \Lambda$$

Let us now consider a perturbation around the previously described solution of the form
\begin{equation}\delta G_{MN}=\delta_M^{\mu}\delta_N^{\nu}h_{\mu\nu}(x,y)\,\,\,\,,\,\,\,\,\delta\phi=0\end{equation}
$h_{\mu\nu}$ represents the graviton in the axial gauge defined by the constraint $h_{5M}=0$. We are interested on the transverse modes and, therefore, we shall assume
\begin{equation}h_{\mu}^{\mu}=\partial_{\mu}h^{\mu\nu}=0\end{equation}
The equations of give, to first order in $h_{\mu\nu}$,
\begin{equation}\left\{-\frac{1}{2}\frac{\partial^2}{\partial y^2}+\left(A^{''}+2(A')^2\right)-\frac{1}{2}e^{-2A(y)}\partial^2\right\}h_{\mu\nu}(x,y)=0\end{equation}
where $\partial^2\equiv \eta^{\mu\nu}\partial_{\mu}\partial_{\nu}$. Introducing a trial solution in the form of a product
of an ordinary-space plane wave times a {\textit{bulk wave function}} $h_{\mu\nu}=e^{ip\cdot x}\psi_{\mu\nu}$, we get the
{\textit{Schroedinger-like}} equation
\begin{equation}\left\{-\frac{1}{2}\frac{d^2}{dy^2}+\left(A^{''}+2(A')^2\right)\right\}\psi(y)
=\frac{m^2}{2}e^{-2A(y)}\psi(y)\end{equation}
We have dropped the spacetime indices and introduced the mass $m^2=-p^2$. The existence of localized graviton in ordinary space amounts to
the existence of a normalizable localized bound state of this equation at zero energy $m^2=0$ (zero-mode). It is not difficult to see that indeed such a zero mode exists. It has the wave-function
\begin{equation}\psi_0(y)=Ne^{2A(y)}\end{equation}
I order to study the massive spectrum we must trasform the above equation into a conventional Schroedinger equation.
This can be done with the help of the transformation
\begin{equation}y\rightarrow z=f(y)\,\,\,\,,\,\,\,\,\,\,\psi(y)=\Lambda \overline{\psi}\end{equation}
 Demanding the absence of first derivative terms and a standard constant coeffisient $m^2$ in the right hand side, we arrive at
\begin{equation}\Lambda =e^{A/2}\,\,\,\,\,, \,\,\,\,\,\,f'=e^{-A}\end{equation}
The Schroedinger equation is
\begin{equation}\left\{-\frac{1}{2}\frac{d^2}{dz^2}+\overline{U}(z)\right\}\overline{\psi}(z)=m^2\overline{\psi}(z)\end{equation}
the potential $\overline{U}(z)$ is
\begin{equation}\overline{U}(z)=\frac{3}{8}e^{2A}\left(2A^{''}+5(A')^2\right)\end{equation}
An alternative expression in terms of derivatives with respect to $z$ is
\begin{equation}\overline{U}(z)=\frac{3}{4}\left(\ddot{A}+\frac{3}{2}(\dot{A})^2\right)\end{equation}
Note that the Schroedinger equation has the form corresponding to supersymmetric Quantum Mechanics
\begin{equation}{\cal{Q}}^{\dagger}{\cal{Q}}\overline{\psi}=\left\{-\frac{d}{dz}-\frac{3}{2}\dot{A}\right\}\left\{\frac{d}{dz}-\frac{3}{2}\dot{A}\right\}\overline{\psi}=2m^2\overline{\psi}\end{equation}
This form clearly excludes the possibility of tachyonic states. Nevertheless, in order to know whether there is a gap in the continuum spectrum we need to know the asymptotic behaviour of the potential. Unfortunately, the change of variable
$z=\int ^ydy\,e^{-A(y)}$ is not analytically integrable. Nevertheless, it is possible to
 argue that since $z$ tends to infinity exponentially, i.e. $\lim_{y\rightarrow \infty}\{z\}\propto e^{\xi^2|y|}$,
\begin{equation}\lim_{z\rightarrow \infty}\left\{\overline{U}(z)\right\}=\lim_{y\rightarrow \infty}\left\{\overline{U}(z(y))\right\}
\propto e^{2A}\rightarrow 0\end{equation}
Therefore, there is no gap and the continuous spectrum starts from zero
energy. Other cases have been discussed in \cite{KGG}

\section{Localization of fermions}

It is know that bounce-like static solutions like $\phi_b$ in eq.(\ref{f})
may trap fermionic zero modes \cite{JR},\cite{CH}.
A bulk fermion coupled to the bounce in
$2n+1$ dimensions gives rise to two chiral fermionic zero modes in
$2n$ dimensions. One of these modes is localized at the bounce and
normalizable while the other in delocalized and non-normalizable.
Thus, chiral fermions may emerge in $2n$ dimensions despite the
fact that they arise from a higher dimensional set up and fermions of both
chiralities would be expected (this is normally the case in dimensional reduction).
The idea of sending the second chiral mode in some extra dimension
is also behind the proposal for solving the doubling problem on
the lattice \cite{Kap}.

Here we will examine if chiral fermionic zero modes still exist in
the presence of gravity The same question has also be discussed in
\cite{DSS} while fermion localization in $AdS$ slices in \cite{GP},\cite{BG}.
For this, we must solve the Dirac equation
\begin{equation}
\Gamma^M\nabla_M\Psi+f\phi\Psi=0\, ,
\end{equation}
where
$\nabla_M=\partial_M+{1\over4}\omega_{ABM}\Gamma^{AB}$ is the
spin-covariant derivative for the background of eq.(\ref{met}) and $\phi$
is the bounce $\phi_B$ in eq.(\ref{f}). The parameter $f$ stands for a, dimensionful in five dimensions, Yukawa coupling.
The only non-vanishing
component of the spin-connection $\omega_{ABM}$ is
$$
\omega_{5\alpha\mu}=-e^{A}A'\eta_{\alpha\mu}\, .
$$
By writing
$$
\Psi(x,y)=\eta(y)\psi(x)\, ,
$$ we get that $\psi$ is a four-dimensional massless fermion
$$
\Gamma^\mu\partial_\mu \psi=0\, ,
$$
for  $\eta$ which satisfies
\begin{eqnarray}
\eta'+2A'\eta+f\phi_B\eta=0\, , &&  \mbox{for}~~~
\Gamma^5\psi=+\psi\, , \\
\eta'+2A'\eta-f\phi_B\eta=0\, , &&  \mbox{for}~~~
\Gamma^5\psi=-\psi\, .
\end{eqnarray}
The solutions to the above equations are
\begin{eqnarray}
\eta=\left(\phantom{{1\over X}}\!\!\!\!\!\!\!\!
\cosh(ay)\right)^{4\beta-fv/a}\, e^{\beta\tanh^2(ay)}\, ,  &&  \mbox{for}~~~
\Gamma^5\psi=+\psi\, , \\
\eta=\left(\phantom{{1\over X}}\!\!\!\!\!\!\!\!
\cosh(ay)\right)^{4\beta+fv/a}\, e^{\beta\tanh^2(ay)}\, ,  &&  \mbox{for}~~~
\Gamma^5\psi=-\psi\, .
\end{eqnarray}
Thus, we see that the negative chirality mode is always
delocalized while the positive one may be localized for
$f>4a\beta/v$.
The existence of the four-dimensional fermion $\psi$ is
connected to the normalizability of $\eta$ which in our case
is written as
\begin{equation}
\int dy \, e^{3A}\, |\eta|^2 < \infty\, ,
\end{equation}
and can be deduced from the bulk five-dimensional Dirac action.
For our solution we get
\begin{equation}
\int dy \left(\phantom{{1\over X}}\!\!\!\!\!\!\!\!
\cosh(ay)\right)^{2\beta\mp fv/a}\, e^{\beta/2\tanh^2(ay)} <
\infty\, , ~~~\mbox{for}~~~
\Gamma^5\psi=\pm\psi\, .
\end{equation}
It is clear then that only the positive chirality mode
$\Gamma^5\psi=+\psi$ can exist for $f>2\beta a/v$ while the
negative chirality mode $\Gamma^5\psi=-\psi$ fails to survive on
the bounce. As a result, even in the presence of gravity, there
always exists a chiral massless fermion on a domain wall formed by
the bounce. We will see in a moment that the bounce can
localize not only the massless graviton and  fermions but gauge
fields as well. In the latter case, one additional field is
needed, the dilaton.

\section{Other fields on the bounce. The dilaton.}

There has been renewed interest in modeling our universe
as a domain wall, motivated by the D-brane solutions of string
theory. A crucial difference of a domain wall to the D-branes
is the fact that the latter have gauge fields living in their
world-volumes. These gauge fields emerge  from open strings
ending on the D-branes. However, in the case of a domain wall,
which as we have seen already can be built from a bounce even in the
presence of gravity, no gauge fields
are expected on the wall. Thus, one should look for other sources
which hopefully will provide the gauge fields. The most obvious
such source is of course the bulk of space-time. That is, bulk
gauge fields may deposit  zero modes on the wall building this
way the four-dimensional gauge sector. On the other hand, bulk
gauge fields are ``claustrophobic" in the sense that they
avoid to be  localized \cite{DSh}. The basic reason is that a
four-dimensional $U(1)$ theory for example is conformal invariant
and thus warp factors as the one appearing in the
 metric (\ref{met}) cannot be seen by the gauge bosons. As a result, localization of
 bulk gauge fields cannot be achieved as long as the bulk gauge
 theory is in a Coulomb phase \cite{PR}. In contrast, for a five-dimensional
 gauge theory in a strong coupling phase, lattice simulations have
 shown \cite{DFKK} that a four-dimensional layer phase exists in
 which a $U(1)$ theory  appears in Coulomb phase on the
 four-dimensional layer-wall. Here, we will show that gauge field
 localization can be driven alternatively by
  another scalar, the dilaton. The dilaton has also been employed for localization in the so-called
  dilatonic-domain wall \cite{DY}.
    As the dilaton couples to the
$F^2$ term, localization is in principle possible. The plan is
therefore, to find a non-singular background solution to the bounce-gravity-dilaton
system. Then, by solving Maxwell
equations on this background, we will examine if a localized
four-dimensional massless photon exists.
For this, let us consider the action
\begin{equation}
{\cal{S}}=\int d^4x\int dy\sqrt{-G}\left\{2M^3R-\frac{1}{2}(\partial \phi)^2
-\frac{1}{2}(\partial \pi)^2 -V(\phi,\pi)\right\}\, , \label{act}
\end{equation}
$\phi$ stands for the scalar field destined to obtain a bounce-like configuration
while $\pi$ is another scalar to which
we loosely give  the name {\textit{dilaton}}. Later we will see,
by choosing an appropriate potential, that $\pi$ is a true
dilaton.
The potential is in general both $\phi$ and $\pi$ dependent.
The equations of motion resulting from (31) are
\begin{equation}R_{MN}-\frac{1}{2}G_{MN}R=
\frac{1}{4M^3}\left\{\partial_M\phi\partial_N\phi+\partial_M\pi\partial_N\pi
-G_{MN}\left(\frac{1}{2}(\partial \phi)^2+
\frac{1}{2}(\partial \pi)^2+V\right)\right\}\, ,
\end{equation}
\begin{equation}
\frac{1}{\sqrt{-G}}\partial_M\left\{\sqrt{-G}G^{MN}\partial_N\chi\right\}
=\frac{\partial V}{\partial \chi}\, ,
\end{equation}
with $\chi=\phi,\,\pi$.
Introducing a metric ansatz of the form
\begin{equation}
G_{MN}=\left(\begin{array}{cc}
e^{2A(y)}\eta_{\mu\nu}&0\\
0&e^{2B(y)}
\end{array}\right)\, ,
\end{equation}
and restricting ourselves to scalar field solutions depending only on the fifth coordinate $y$,
leads us to the set of equations
\begin{equation}
\frac{1}{2}(\phi')^2+\frac{1}{2}(\pi')^2-e^{2B}V=24M^3(A')^2\, ,
\end{equation}
\begin{equation}
\frac{1}{2}(\phi')^2+\frac{1}{2}(\pi')^2+e^{2B}V
=-12M^3A^{''}-24M^3(A')^2+12M^3A'B'\, ,
\end{equation}
\begin{equation}
\chi^{''}+\left(4A'-B'\right)\chi'
=\partial_{\chi}V\,\,,\,\,\,\,\,\chi=\phi,\pi\, ,
\end{equation}
out of which three are independent.
Solutions to the above system of equations can be
found by following the {\textit{superpotential method}}\cite{G} for which $\phi'=\frac{\partial{\cal{W}}}{\partial \phi}$.
The particular solution we will consider here corresponds to the
potential
\begin{equation}
V=e^{
\pi/\sqrt{12M^3}}
\left\{\frac{1}{2}\left(\frac{\partial {\cal{W(\phi)}}}{\partial \phi}\right)^2-\frac{5}{32M^3}{\cal{W(\phi)}}^2\right\}
\end{equation}
whereas the solution is
\begin{equation}
\pi=-\sqrt{3M^3} A\,,\,\,\,\, B=\frac{A}{4}=-\frac{\pi}{4\sqrt{3M^3}} \,,\,\,\,\,A'=-{1\over 12M^3} W\, . \label{bb}
\end{equation}
Thus, by choosing a specific $W(\phi)$, the solution is completely
specified. For example, the superpotential in eq.(\ref{w}),
gives
\begin{eqnarray}
&&\phi(y)=v\tanh(ay)\, , ~~~
A(y)=-\beta \ln \cosh^2(ay)-\frac{\beta}{2}\tanh^2(ay)\, , \label{aa}\\
&& \pi(y)=\frac{\beta}{\sqrt{3M^3}}\left( \ln
 \cosh^2(ay)+\frac{1}{2}\tanh^2(ay)\right)\, ,
\end{eqnarray}
 where
$ \beta=v^2/ 36M^3$.

It is not difficult now to see that the geometry in
eq.(41) with $A(y)$ given in eq.(\ref{aa}) is, contrary to
what advertised, singular. However, this singularity is of the
same nature as the singularities encountered in various D-brane
solutions is string theory. For example a D4-brane in type IIA
theory is singular since it has a diverging dilaton field (our
$\pi$). This singularity, however, disappears when the D4 is
lifted to eleven-dimensional supergravity and the dilaton turns to
be the radius of the eleventh compactified dimension. Trying the
same in our solution, we arrive at the same result. The solution
can be lifted to six dimensions as
\begin{eqnarray}
&&
ds_6^2=e^{3A(y)/2}\left(-dt^2+dx_1^2+dx_2^2+dx_3^2+dz^2\right)+dy^2
\end{eqnarray}
where $z$ parametrize an extra $S^1$ direction. String-like solutions in six-dimensions
with localized graviton have been constructed in \cite{GSS}. The metric above
is everywhere regular for the bounce eq.(\ref{aa}). The fact that
we can lift the five-dimensional solution to six dimensions where
$\pi$ is just the radius of the extra dimension, justifies the
name dilaton for $\pi$. It should be noted here that similarly,
the self-tuning solution \cite{KS} which is singular in five dimensions can
also be lifted to six dimensions. The metric one finds in this
case has been given in \cite{AK} and it is non-singular.

Let us complete this section by considering graviton localization
 in the above background $ds^2=e^{2A(y)}\eta_{\mu\nu}dx^{\mu}dx^{\nu}+e^{2B(y)}dy^2$. The standard
perturbation $\delta G_{MN}=\delta_{M}^{\mu}\delta_{N}^{\nu}h_{\mu\nu}(x,y)$ leads to the graviton equation\footnote{Again
$h_{5M}=h_{\mu}^{\mu}=\partial_{\nu}h_{\mu}^{\nu}=0$ and $h_{\mu\nu}(x,y)=e^{ip\cdot x}\psi_{\mu\nu}(y)$.}
\begin{equation}-\psi^{''}(y)+B'\psi'(y)+2\left(A^{''}-A'B'+2(A')^2\right)\psi(y)=m^2e^{2(B-A)}\psi(y)\end{equation}
Our solution has $B=A/4$. For $m=0$ it is easy to see that we have a normalizable zero-mode
$$\psi_0(y)=e^{2A(y)}$$
The massive graviton spectrum can be probed with the help of the change of variables
$$z=\int dy\,e^{-3A(y)/4}\,\,,\,\,\,\,\psi(y)=e^{A(y)/2}\eta(z)$$
The resulting equation
\begin{equation}-\ddot{\eta}+\tilde{V}\eta=m^2\eta\end{equation}
contains a potential $\tilde{V}=\frac{3}{2}e^{3A/2}\left(A^{''}+\frac{9}{4}(A')^2\right)=\frac{3}{2}\left(\ddot{A}+\frac{3}{2}(\dot{A})^2\right)$ that
vanishes at infinity. It can also be cast in a supersymmetric Quantum Mechanics form
$-\left(\frac{d}{dz}+\frac{3}{2}\dot{A}\right)\left(\frac{d}{dz}-\frac{3}{2}\dot{A}\right)\eta=m^2\eta$.
Therefore, the spectrum is gapless and without tachyons.

\section{Localization of gauge bosons}

Normally, as we have already noticed, idependently of the metric function,
no localization is possible due to the rescaling properties of the gauge boson
kinetic term that make the warp factor drop out.
Nevertheless, the dependence of the gauge coupling
on the dilaton can change that.
 When gauge bosons are included, the action eq.(\ref{act}) turns
 out to be
\begin{equation}
{\cal{S}}=\int d^4x\int dy\sqrt{-G}\left\{2M^3R-\frac{1}{2}(\partial \phi)^2
-\frac{1}{2}(\partial \pi)^2 -V(\phi,\pi)-{1\over 4g^2}e^{-\lambda \pi/2\sqrt{3M^3}}F_{MN}F^{MN}\right\}\, ,
\end{equation}
where $\lambda$ is a dimensionless dilaton coupling depending on the details of the underlying theory. We will
examine for simplicitly the case of an abelian $U(1)$ theory. In
this case, the  equations of motion turns out to be
\begin{equation}
{1\over
\sqrt{-G}}\partial_M\left(\sqrt{-G}G^{MK}G^{NL}e^{-\lambda\pi/2}F_{KL}\right)=0\, .
\label{ff}
\end{equation}
In the gauge
\begin{equation}
\partial_{\mu}A^{\mu}=A_5=0\, ,
\end{equation}
and for the background $ds^2=e^{2A(y)}\eta_{\mu\nu}dx^{\mu}dx^{\nu}+e^{2B(y)}dy^2$,
the equations of motion take the form
\begin{equation}
\partial^2A_{\mu}+e^{-B}e^{\lambda \pi/2\sqrt{3M^3}}\frac{\partial}{\partial y}
\left\{e^{2A-B}e^{-\lambda \pi/2\sqrt{3M^3}}\frac{\partial}{\partial
y}A_{\mu}\right\}=0\, .
\end{equation}
For an ansatz
\begin{equation}
A_{\mu}(x,y)=u^{\mu}(x)U(y)=u^{\mu}(0)e^{ip\cdot x}U(y)
\end{equation}
with $p^2=-m^2$, we obtain the equation
\begin{equation}-\frac{d^2}{dy^2}U(y)-\left(2A'(y)-B'(y)-
\lambda \pi'(y)/2\sqrt{3M^3}\right)\frac{d}{dy}U(y)=\frac{m^2}{2}e^{2(B-A)}U(y)\, .
\end{equation}
It is clear that for $m^2=0$, the above equation has a constant (zero-mode) solution
\begin{equation}
U_0=constant\, .
\end{equation}
In the absence of the dilaton, this zero mode is non-normalizable
and thus, there is no localizable massless photon on the bounce.
However here, thanks to the presence of the dilaton, the gauge vector
constant zero-mode is now localizable, just as a scalar constant zero-mode is.
Indeed, we have
\begin{equation}
\Delta {\cal{S}}=-\frac{1}{4g^2}\int d^4x f_{\mu\nu}(x)f_{\alpha\beta}(x)
\eta^{\mu\alpha}\eta^{\nu\beta}\int_{-\infty}^{+\infty}
dy \,U_0^2e^{B(y)-\lambda \pi(y)/2}\end{equation}
According to the solution found $B=-\pi/4\sqrt{3M^3}$. The last integral is, of course, finite
$$\int_{-\infty}^{+\infty}dy\,e^{-\frac{(1+2\lambda)}{4}\frac{\pi(y)}{\sqrt{3M^3}}}=
\int_{-\infty}^{+\infty}dy\,e^{(1+2\lambda)A(y)/4}=$$
$$\int_{-\infty}^{+\infty}dy\left(\cosh(ay)\right)^{-\beta(1+2\lambda)/4}e^{-\beta(1+2\lambda)\tanh^2(ay)/8}<\infty$$
It should be noted that the gauge vector localization is a strong-coupling
property in agreement with the increase of the gauge coupling
away from the {\textit{brane}}
$$g_{eff}^2\propto g^2
\left(\cosh(ay)\right)^{\beta(1+2\lambda)/4}e^{\beta(1+2\lambda)\tanh^2(ay)/8}$$
As far as the massive vector spectrum is concerned we can consider a transformation
$$z=\int^y dy\,e^{-3A(y)/8}\,\,\,\,\,\,,\,\,\,\,\,U=e^{-\zeta A}\Omega(z)$$
with $\zeta=(11+4\lambda)/16$, and transform the above wave-equation into a standard Schroedinger form
\begin{equation}\left\{-\frac{d^2}{dz^2}+\tilde{V}\right\}\Omega(z)=m^2\Omega(z)\end{equation}
with $\tilde{V}= \zeta e^{3A/4}\left(A^{''}+(3/8-\zeta)(A')^2\right)=\zeta \left(\ddot{A}+\zeta(\dot{A})^2\right)$.
The exponential factor forces the limit $\lim_{z\rightarrow \infty}\left\{\tilde{V}(z)\right\}=0$ that implies no gap. Note also that, again, the supersymmetric form of the Schroedinger operator
$\left\{\frac{d}{dz}-\zeta\dot{A}\right\}\left\{\frac{d}{dz}+\zeta\dot{A}\right\}\Omega=-m^2\Omega$.

\section{Dilaton-induced localization on a brane}

It should be stressed that the localization properties of the gravity-dilaton system are not particular to the smooth spaces
defined in terms of the bounce but hold also in the case that we start off with a brane. Let us start with the action
\begin{equation}
{\cal{S}}=\int d^4x\int dy\sqrt{-G}\left\{2M^3R-\frac{1}{2}(\partial \pi)^2-V(\pi)\right\}
-\vartheta\int d^4x \sqrt{-\det{G_{\mu\nu}(x,y=0)}}\end{equation}
where $\vartheta$ stands for the brane tension. The resulting equations of motion are
\begin{equation}R_{MN}-\frac{1}{2}G_{MN}R=\frac{1}{4M^3}\left\{\partial_M\pi\partial_N\pi-G_{MN}\left(\frac{1}{2}(\partial \pi)^2+V\right)-G_{\mu\nu}\delta_M^{\mu}\delta_N^{\nu}\vartheta \delta(y)\right\}\end{equation}
\begin{equation}\frac{1}{\sqrt{-G}}\partial_M\left\{\sqrt{-G}G^{MN}\partial_N\pi\right\}=\frac{\partial V}{\partial \pi}\end{equation}
For the metric ansatz
$$G_{MN}=\left(\begin{array}{cc}
e^{2A(y)}\eta_{\mu\nu}\,&\,0\\
0\,&\,e^{A(y)/2}
\end{array}\right)$$
 and for $\pi$ depending only on the fifth dimension, these equations reduce to the pair of independent equations
\begin{equation}(A')^2=\frac{1}{24M^3}\left(\frac{1}{2}(\pi')^2-V(\pi)e^{A/2}\right)\end{equation}
\begin{equation}A^{''}+\frac{7}{4}(A')^2=-\frac{1}{12M^3}\left(\frac{1}{2}(\pi')^2+V(\pi)e^{A/2}\right)-\frac{\vartheta}{12M^3}\delta(y)\end{equation}
The parameter $\vartheta$ stands for the brane tension. These equations are solved with a Randall-Sundrum warp function
$$A(y)=-\frac{\vartheta}{24M^3}|y|\,\,,\,\,\,\,\pi(y)=-A(y)\sqrt{3M^3}=\frac{\vartheta}{8\sqrt{3M^3}}|y|$$
and for the dilaton potential choice
$$V(\pi)=-\frac{5\vartheta^2}{128M^3}e^{\pi/2\sqrt{3M^3}}$$

Gauge bosons can be localized through their dilaton coupling in the same fashion as in the bounce case. A constant gauge boson wave-function $U_0$ is a normalizable zero mode. Again, we have
$$\Delta {\cal{S}}=-\frac{1}{4g^2}\int d^4x f_{\mu\nu}(x)f_{\alpha\beta}(x)
\eta^{\mu\alpha}\eta^{\nu\beta}\int_{-\infty}^{+\infty}
dy \,U_0^2e^{B(y)-\lambda \pi(y)/2}$$
The last integral is now $\int_{-\infty}^{+\infty}e^{-\vartheta(1+2\lambda)|y|/96M^3}=192M^3/(1+2\lambda)\vartheta$.
It should be noted that also in this case the solution can be lifted to six
dimensions.

\section{Conclusions}

Summarizing the subject matter of the paper, we have
considered smooth
 spaces, which have
 four-dimensional Poincar\'e invariance, with scalar fields realizing bounce-like solutions.
It is known that solitons like domain walls or strings like the bounce we have considered,
 may support massless fields \cite{JR},\cite{CH},\cite{LL}. In theories
with domain walls formed by a scalar field (bounce)
for example, scalars as well as fermions with appropriate
interactions in the bulk give rise to massless scalars and chiral fermions
on the wall \cite{LL},\cite{JR}. Fermions in five-dimensions in particular,
with Yukawa couplings to the bounce, deposit a single chiral fermionic
zero mode on the four-dimensional wall.  We have seen here that
when the scalar field couples to gravity, the coupling of the
bounce to the bulk fermions still leads to massless chiral modes
on the wall. In addition, a massless graviton is localized on the bounce
leading to four-dimensional gravity.

Although graviton and fermions can easily be localized, there is no fully
satisfactory localization
mechanism for gauge fields, although there exists lattice evidence that this is indeed
the case \cite{DFKK}, \cite{DFK}, at least in a Randall-Sundrum background. The reason, as it has been mentioned
previously, is that the action of the gauge fields cannot see the
warp factor of the geometry due to scale invariance in
four-dimensions. The way this is overcome here and made possible
the localization of the gauge field is by the introduction of a
second scalar, the dilaton. As the latter is coupled to the
kinetic term of the gauge fields, localization may be achieved. In
our case, the dilaton coupling to the gauge and scalar fields was
such that the theory can be lifted to six-dimensions with a
five-dimensional $S^1$-compactified wall.

As a final comment, let us note that the same discussion is not
restricted to bounce-like solutions in five or
six dimensions. The same can be done in principle in any
dimension. Particularly interesting is the case of a bounce like
solution in seven dimensions. Given the existence of the chiral
zero modes on the bounce, the six-dimensional theory on the bounce
suffers from gravitational anomalies. In order to cancel the anomalies extra gauge fields are needed on the bounce.
This situation is the same with the Horava-Witten picture \cite{HW}.
There, the theory on walls at the orbifold points of the $S^1/Z_2$
compactification of the eleven-dimensional M-theory, has
gravitational anomalies. The latter are cancelled by appropriate
gauge anomalies. The cancellation specifies the gauge group to be
$E_8$ at each one of the walls. However, in the six-dimensional
bounce, there is no unique specification of the gauge group
since the gravitational anomalies can be cancelled by many choices of the latter.
\vspace{.9cm}

\noindent
{\textbf{Acknowledgements}}

The authors would like to thank the CERN Theory Division for
hospitality. K. T. acknowledges traveling support from the TMR
program ``Beyond the Standard Model". This work is partially
supported by the  RTN contracts HPRN-CT-2000-00122 and
HPRN-CT-2000-00131 and the $\Gamma\Gamma$ET grant E$\Lambda$/71.

\end{document}